# Temporal Variation Measure Analysis: An Improved Second-Order Difference Plot

Chen Diao[1,2,*], Ning Cai[3]

1. School of Information Engineering, Ningxia University, Yinchuan 750021, China
2. Collaborative Innovation Center for Ningxia Big Data and Artificial Intelligence Co-founded by Ningxia Municipality and Ministry of Education, Ningxia University, Yinchuan 750021, China
3. School of Artificial Intelligence, Beijing University of Posts and Telecommunications, Beijing100876, China

Correspondence should be addressed to Chen Diao; diaochen@nxu.edu.cn

**Abstract:** In this study, an improved second-order difference plot is proposed to analyze the variability of heart rate variability. Although the variation of physiological status of cardiovascular system can be shown graphically by the second-order difference plot, the descriptive ability of existing indicators for this plot is insufficient. As a result, the physiological information contained in the second-order difference plot cannot be extracted adequately. Addressing the problem, the temporal variation measure analysis is presented to describe distribution patterns of scatter points in the second-order difference plot quantitatively and extract the acceleration information for variation of heart rate variability. Experiment results demonstrate the effectiveness of the temporal variation measure analysis. As a quantitative indicator, the temporal variation entropy is properly designed and successfully applied in the recognition and classification of the physiological statuses of the heart.

# 1. Introduction

Electrocardiogram (ECG) records the abundant physiological information of the heart, which has been widely applied in diagnosis for a variety of cardiovascular diseases. The research of analytical method of ECG is always a hot spot in the field of biomedical signal processing [1-5]. The nonlinear analyses are usually employed to analyze ECG, since it reflects nonlinear characteristic of biomedical signal. In numerous nonlinear analytical algorithms, the Poincaré plot analysis and the second-order difference plot analysis (SODP) are classical and often employed in the research of heart rate variability (HRV) [6-16].

The theoretical foundations of Poincaré plot and SODP is state space reconstruction [17]. Based on the theory, the motion state of nonlinear system can be well reconstructed and revealed in a reconstructed state space. The system performance can further be adequately analyzed via reconstructed trajectories. Poincaré plot is a two-dimensional space constructed by a time series. The plot has the function to graphically reveal the physiological status of cardiovascular system. Based on Poincaré plot, HRV can be effectively analyzed and various cardiovascular diseases display distinct distribution patterns [18-20], for instance, the heart rate of healthy human appears "comet-shape", the patient with serious heart failure is "fan-shaped", and the heart rate produced by artificial pacemaker is "round". For describing these distribution patterns of scatter points quantitatively, many analysis algorithms were presented, such as the minor and major axis of fitted ellipse (SD1, SD2) [21], the fitted ellipse area of scatter points (S) [22], Ehler index (EI) [23], Guzik index (GI) [24], Porta index (PI) [25], and the local distribution entropy [6]. Numerical quantitative indicators have been proposed during the past decade, making Poincaré plot analysis widely applied in biomedical signal analysis.

As an extension of the Poincaré plot, Cohen *et al.* [26] presented the second-order difference plot and central tendency measure (CTM) based on the chaos theory in 1996, which were firstly used to quantify the variability of physiological status of the heart. As SODP is constructed by the second-order difference of successive RR intervals, the variation of RR intervals can be reflected in a two-dimensional space graphically with various distribution patterns of the reconstructed data points representing different physiological statuses of cardiovascular system. Although a great deal of information on the variability of HRV is contained in SODP, a bit of information is hard to be adequately extracted by existing quantitative indicators yet. These indicators only depict the overall distribution of points in SODP, whereas the variation tendency for successive RR intervals contained in a single point is ignored natively.

The motivation of study comes from existing limitations of SODP. We mainly focus on the relationship between the position of reconstructed scatter point in SODP and the increment of variation of RR intervals. The information of increment reflects the variation tendency of heart rate within a short time and helps us analyze physiological status of heart more comprehensively. On the basis of our previous researches [6, 27-30], we design an improved second-order difference plot in this study and a more effective feature of physiological status of heart can be extracted via the novel algorithm. The main contributions of this paper are as follows:

1. Temporal variation measure algorithm (TVM) is presented in this paper. Via TVM, both the information of the variation and the variation tendency for successive RR intervals can be fused and graphically reflected in a three-dimensional second-order difference plot.
2. For TVM, the temporal variation entropy is proposed, which can assess the distribution of points in the three-dimensional second-order difference plot more objectively.

The arrangement of this paper is organized as follows. SODP are briefly introduced in Section 2. Section 3 provides the theory and steps of TVM in detail. In Section 4, experiments are given to demonstrate the performance of our method. Finally, the major outcomes are given in Section 5.

## 2. Second-order Difference Plot Analysis

### 2.1. Second-order Difference Plot

As a significant extension of the Poincaré plot analysis, SODP focuses primarily on describing the variability of successive RR intervals. Here, SODP is computed by plotting $Y(n)$ against $X(n)$ which are given by equations (1) and (2), respectively.

$$X(n) = x(n+1) - x(n) \tag{1}$$

$$Y(n) = x(n+2) - x(n+1) \tag{2}$$

where $x(i)$ is the $i$th time interval of successive RR interval series.

Based on theory of state reconstruction, the coordinates $(X(i), Y(i))$ of all points can be calculated and the distribution pattern of all scatter points constructed by the time series is visually shown in SODP.

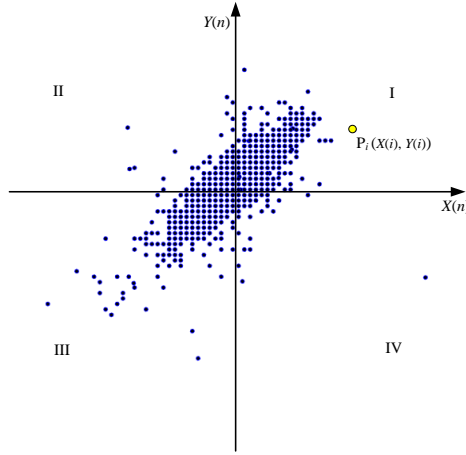

Figure 1. SODP of a normal human.

Figure 1 illustrates the SODP of a normal human. In the plot, the coordinates of point $P_i$ are composed of three successive RR intervals. The horizontal and vertical coordinates of this point are $x(i+1)-x(i)$ and $x(i+2)-x(i+1)$ respectively. In the plot, the points are located in different quadrants representing the diverse physiological statuses of heart rate. For the point $P_i$ in the first quadrant (I), the coordinate values of $X(i)$ and $Y(i)$ are greater than zero, which means that $x(i+2) > x(i+1) > x(i)$. Thus, these points in the first quadrant indicate that successive RR intervals increase monotonically and the heart rate decrease steadily. Oppositely, the points in the third quadrant (III) imply increasing heart rate. If the changes of three successive RR intervals are inconsistent, the corresponding points will be located in the second quadrant (II) or the fourth quadrant (IV). Therefore, SODP can describe the variability of heart rate intuitively.

## 2.2. Central Tendency Measure

As a matched indicator of SODP, the central tendency measure (CTM) can quantify the degree of variability of the heart rate. In SODP, CTM is a ratio of the number of points in radius $r$ to the number of all points. The indicator is calculated as follows:

$$CTM(r) = \frac{\sum_{i=1}^{n-2} \delta(d_i)}{n-2} \qquad (3)$$

$$\delta(d_i) = \begin{cases} 1 & if\ ([x(i+2)-x(i+1)]^2 + [x(i+1)-x(i)]^2)^{0.5} < r \\ 0 & otherwise. \end{cases} \qquad (4)$$

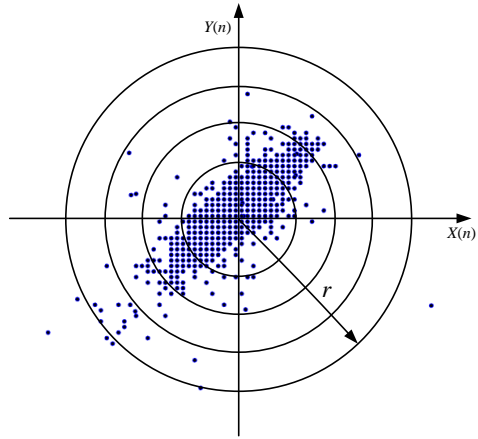

Figure 2. Analysis of SODP with radius $r$.

Figure 2 shows that CTM counts the number of all points in the origin centered circles with radius $r$. As the quantitative feature of distribution density for scatter points, the parameter $r$ decides the value of CTM directly, which influence the classification accuracy significantly. Nevertheless, how to reasonably choose an optimal radius $r$ is still problematic.

SODP describes the variability of successive RR intervals graphically. The scatter points in different quadrants of SODP contain abundant physiological information. Even some points are located in a certain quadrant, there are still much information of cardiovascular system waiting to be analyzed further.

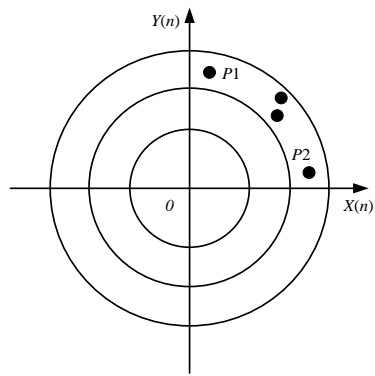

Figure 3. The limitations of CTM.

In Figure 3, four points are located in the same area according to the algorithm of CTM. Hence, the four points will be counted indiscriminately and used to calculate CTM. However, the four points represent the different variation tendencies of the RR interval series, which cannot be distinguished by CTM effectively. For example, as the large coordinate difference of the point $P1$, the point is located next to the y-axis and the position of *P*1 means different variation for three successive RR intervals is an incremental tendency.

Conversely, the point *P*2 near the x-axis reflects different variation of successive time intervals is a diminishing tendency. Therefore, CTM cannot comprehensively describe the variation of the RR intervals. Addressing the problem, an improved second-order difference plot analysis is presented in the next section, which will improve the performance of SODP further.

## 3. Temporal Variation Measure Algorithm

For the limitations of SODP, in this section, the temporal variation measure algorithm is introduced in detail, which is an improvement for standard SODP. The algorithm is composed of three dimensional second-order difference plot and temporal variation entropy. In Section 3.1, the RR interval series is reconstructed in a three dimensional second-order difference plot. In the space, the distribution patterns of reconstructed scatter points reflect more physiological information of cardiovascular system. Then, for analyzing these patterns quantitatively, a temporal variation entropy is designed in Section 3.2.

## 3.1. Three Dimensional Second-order Difference Plot

Rich physiological information of the heart is shown by the distribution pattern of scatter points. In SODP, the position and Euclidean distance of points reflect the variation tendency of the RR intervals essentially, which can be explained by the concept of acceleration. However, the acceleration of the variation for the RR intervals cannot be reflected by the indicators of CTM effectively. For this limitation, the temporal variation measure algorithm is presented. Based on the novel analysis method, the change and the acceleration of the change for the RR intervals will be calculated adequately.

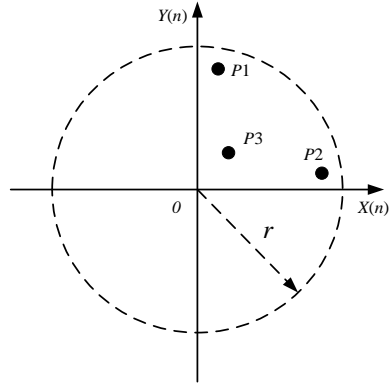

Figure 4 The relationship between the position of point and the variability of RR intervals.

In Figure 4, there are three points *P*1, *P*2 and *P*3 in SODP. As mentioned above, the distance between the points *P*1 and the origin is the same as the distance between the *P*2 and the origin, whereas the two points reflect the different conditions of the heart. On the other hand, the different distances also represent the different physiological statuses of the heart. For instance, the distance between *P*3 and the origin indicates that the changes of three successive RR intervals are smaller than the other two points'. Hence, the two factors of position and distance of points should be properly fused in the novel algorithm.

Various heart rate accelerations can be described by the different positions of scatter points in SODP. Those scatter points near the vertical axis of SODP reflect that the accelerations of heart rates are increasing. On the other hand, if the points are located at the abscissa axis of SODP nearby, this situation shows that the accelerations of variation for the heart rate are reducing. Hence, the relative positions between the points and coordinate axes can describe the acceleration of variation for heart rate graphically. Here, the relative position of point in SODP can be well calculated by the parameter $D_{co}$.

$$D_{co}(i) = |x(i+2) - x(i+1)| - |x(i+1) - x(i)| \tag{5}$$

where $D_{co}(i)$ is the difference of vertical and horizontal coordinates of point $P_i$. If the parameter $D_{co}$ is greater than zero and the point is closer to the y-axis, which means that the variation tendency of RR intervals is rising in a short time. Otherwise, the point is near the x-axis and the variation tendency decreases. In fact, the parameter $D_{co}$ provides the acceleration information of variation for RR intervals and well depicts the laws of the variation of the heart conditions.

On the other hand, the variation of RR intervals can be described by Euclidean distance between the point and the origin of SODP, and the RR intervals is positively associated with the distance. the distance can be calculated by the parameter $L$.

$$LE(i) = \sqrt{x_i^2 + y_i^2} \tag{6}$$

$$L(i) = [1 + \exp(-\frac{LE_i}{\overline{LE}})]^{-1} \tag{7}$$

where $LE(i)$ is Euclidean distance between point $P_i$ and the origin. $x_i$ and $y_i$ are the coordinate values of $P_i$ respectively. $\overline{LE}$ is the mean of all $LE$s. Here, formula (7) is the sigmoid function, which can scale the distance $LE$. Via the nonlinear function, the Euclidean distance $LE$ is mapped to the interval (0,1).

For fusing the two factors further and reserving the characteristics of SODP, we design a novel coordinate $z_{TVM}$ for standard SODP, which is calculated as:

$$z_{TVM}(i) = D_{co}(i) \times L(i) \tag{8}$$

In the formulas, two parameters $D_{co}$ and $L$ are fused in $z_{TVM}$, which can describe the variation and the acceleration of variation for RR intervals quantitatively. Hence, the coordinates of points are composed of three coordinates x, y and $z_{TVM}$. In other words, the RR interval time series will be reconstructed in a three-dimensional second-order difference plot. In the novel space, the change of the RR interval series will be shown more comprehensively.

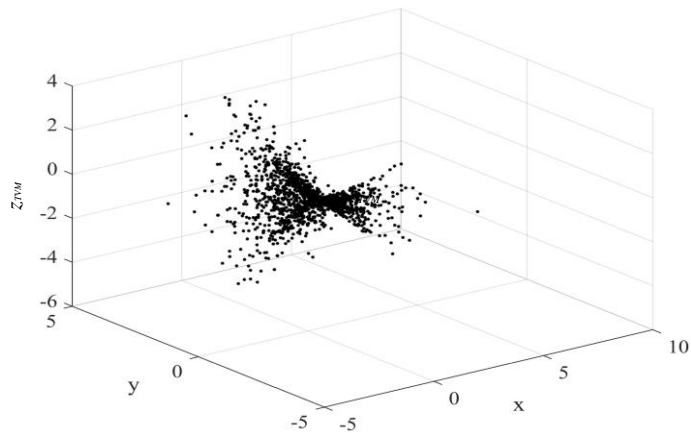

Figure 5. Three dimensional second-order difference plot of normal sinus rhythm RR intervals.

Figure 5 shows a three-dimensional second-order difference plot. The x-y plane is standard SODP in which the projections of scatter points contain all distribution patterns of standard SODP. Meanwhile, the $z_{TVM}$ coordinate of scatter point provides more detailed variation tendency information of the time series, which makes up for the deficiency of SODP. Therefore, these scatter points in the high dimensional SODP can describe the variation of RR interval time series adequately.

## 3.2. Temporal Variation Entropy

The scatter points are fully distributed in the three-dimensional space. Compared with standard SODP, there are more distribution patterns shown in the high dimensional space. For extracting the distribution information of the three-dimensional SODP quantitatively, a quantitative indicator is designed. Inspired by the concept of the local distribution entropy [6], the temporal variation entropy is presented in this study.

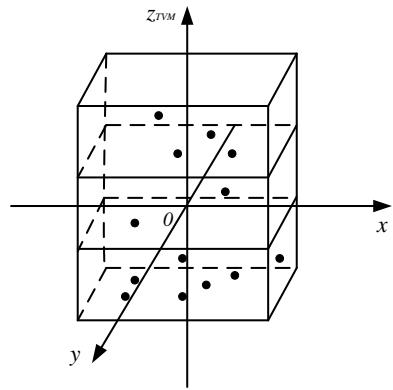

Figure 6. The concept of temporal variation entropy.

In Figure 6, for the three-dimensional SODP, the x-y plane is standard SODP in nature. The axis coordinate $z_{TVM}$ is perpendicular to the x-y plane. In the figure, there is a closed cubic space. The height of the space jointly depends on the maximum value and the minimum value of the $z_{TVM}$ coordinates of all points. The cross section of this space is a rectangle. The maximum value and minimum value of x-coordinate of all scatter points are chosen and the difference of two extreme values is calculated as the length of the cross section. Similarly, the extreme values of y-coordinate can be selected from all scatter points and we will obtain the width of cross section.

Here, the space is divided into several subspaces appropriately. In the different subspace, these scatter points well reserve the distribution information of SODP. Meanwhile, various change tendencies are reflected by these subspaces with different heights. Therefore, we can obtain the distribution information of scatter points by counting the number of the points in every subspace. Here, the concept of entropy is employed to measure information. The temporal variation entropy is defined as follows.

$$E_{TV} = \sum_{i=1}^{N} \sum_{j=1}^{n_i} -n_i |z_{TVM}(j)| \frac{|n_i - \bar{m}|}{M} \log(\frac{|n_i - \bar{m}|}{M}) \qquad (9)$$

where $N$ is the number of the subspace, $n_i$ is the number of the scatter points in the $i$th subspace, $z_{TVM}(j)$ is the coordinate $z_{TVM}$ of the point in the $j$th subspace, $M$ is the number of all points, and $\bar{m}$ is the average number of points. To emphasize the quantity of scatter points and the difference of positions for these points, two weighting factors $n_i$ and $z_{TVM}$ are added in the formula (9). Thus, the variation of RR interval series and the distribution pattern of scatter points can be fully described by $E_{TV}$.

Here the steps of the temporal variation measure algorithm are listed as follows:

(1) calculate the positions of points in SODP by equation (5);

(2) compute the distance between the point and the origin of SODP by equations (6) and (7);

(3) fuse the position parameter $D_{co}$ and the distance parameter $L$, and calculate the coordinate $z_{TVM}$ by equation (8);

(4) select the proper number of subspaces $N$ and calculate the temporal variation entropy $E_{TV}$ by equation (9).

In virtue of TVM, the variation of RR interval time series is adequately revealed in the novel space. Meanwhile, the characteristics of the time series can be quantified by the indicator $E_{TV}$ accurately. To assess the performance of TVM objectively, the algorithm will be applied in the classification of cardiovascular diseases in the next section. Then, the effectiveness of the method will be fully demonstrated by experimental results.

## 4. Application of the Temporal Variation Measure Algorithm

To verify the performance of TVM, in this section, it is applied to extract the features from the normal sinus rhythm and four cardiovascular diseases. In this experiment, five datasets are chosen from PhysioNet/PhysioBank, which are Normal Sinus Rhythm RR Interval Database (nsr2db), CU Ventricular Tachyarrhythmia

- 10 -

Database (cudb) [31], MIT-BIH Arrhythmia Database (mitdb) [32], Post-lctal Heart Rate Oscillations in Partial Epilepsy (szdb) [33] and Smart Health for Assessing the Risk of Events via ECG Database (shareedb) [31], respectively.

In order to demonstrate the performance of TVM, here three indicators, CTM, component CTM (CCTM) and mean distance of the points within the circular radius (D) [34], are tested experiment. Here the quantitative indictor D is the mean distance of these points in a circular radius *r*. The other indictor CCTM is the development of CTM. Based on CCTM, the number of the points in the four quadrants of SODP are counted respectively and the CTM of each quadrant is calculated.

$$CCTM_k(r) = \frac{\sum_{i=1}^{n-2} \delta(d_k(i))}{n-2}, \quad k=1,2,3,4 \tag{10}$$

where *n* is the number of all points, $\delta(d_k(i))$ is the number of points in the *k*th quadrant of SODP.

All three indictors are derived from radius *r*. Thus, how to choose an optimum *r* is a crucial procedure, which can influence the results of experiment directly. Figure 7 shows the values of CTM(*r*) and D(*r*) for five datasets, nsr2db, cudb, mitdb, szdb and shareedb. In figure (a), experiment results of CTM(*r*) indicate that there is distinct difference for these datasets when the radius *r* is greater than 3. Therefore, in the experiment, parameter *r* is set to 3 for the indicator CTM(*r*). Similarly, radius *r* is set to 6 for the indicator D(*r*) in figure (b).

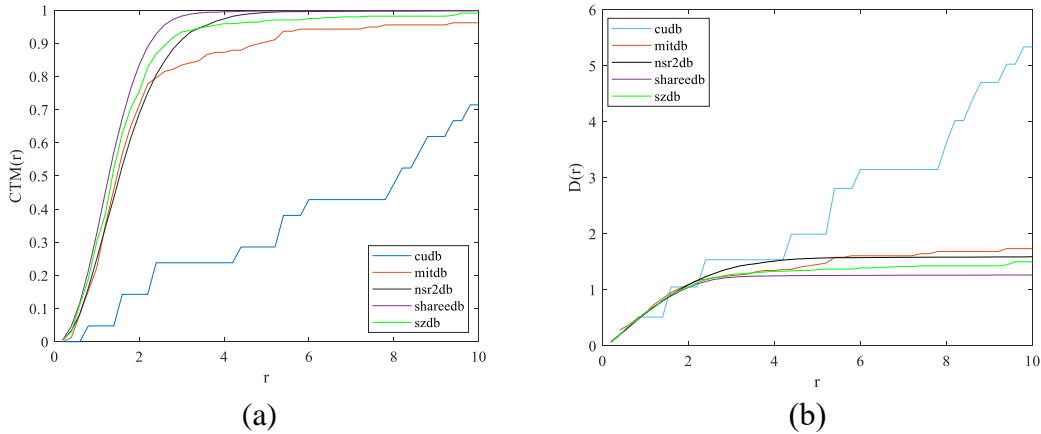

(a)　　　　　　　　　　　　　　(b)

Figure 7. CTM(*r*) and D(*r*) for cudb, mitdb, nsr2db, shareedb and szdb.

The values of CCTM(*r*) in four quadrants are displayed in Figure 8. Figure (a) shows the values of CCTM1(*r*) in the first quadrant, which indicate substantial difference when the $r \in [2, 5]$. Figures (b)~(d) are CCTM2($r$), CCTM3($r$) and CCTM4($r$) respectively. From the three figures, we can find that once the value $r$ is beyond 3, prominent difference would appear for CCTM$_k$($r$). Thus, for CCTM, the parameter $r$ is set to 3 in the experiment.

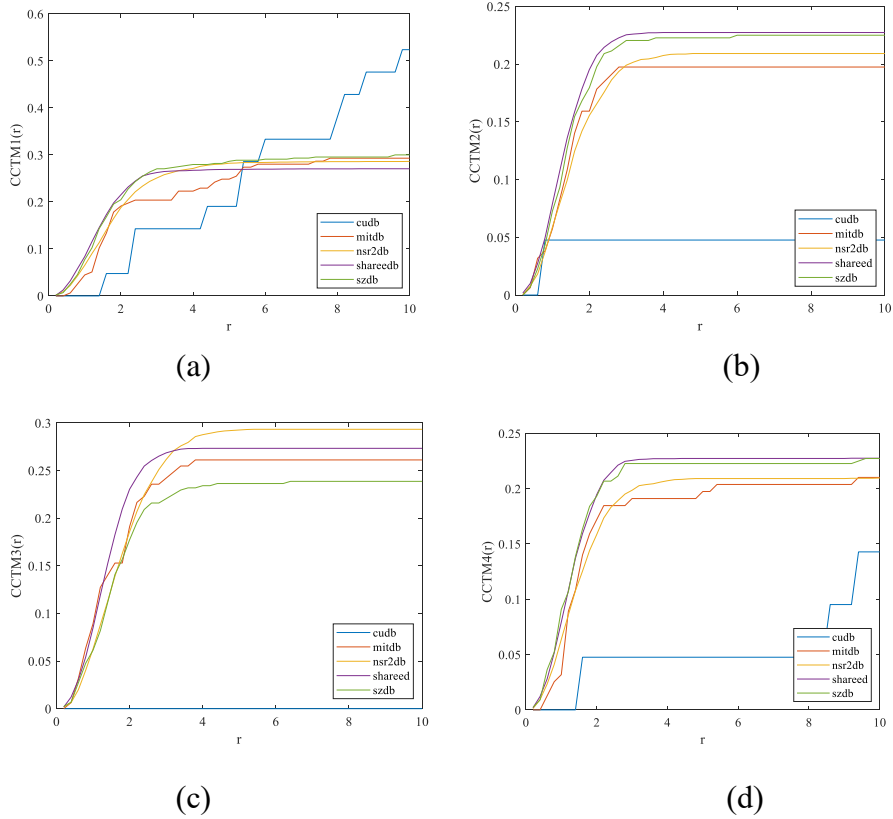

Figure 8. CCTM$_k$($r$) for cudb, mitdb, nsr2db, shareedb and szdb.

Here, the quantitative characteristics of the five datasets are calculated by the indicators, CTM, CCTM and D. Figure 9(a) ~ (f) shows the boxplots of experiment results for CTM, D and CCTM$_k$ (k = 1,2,3,4), respectively.

For the indicator CTM, the numerical ranges of results for nsr2db, shareedb and szdb are shown in figure (a). Meanwhile, in table 1, the means and standard deviations of nsr2db, shareedb and szdb are 0.93±0.01, 0.98±0.01 and 0.94±0.02, respectively. From the boxplots and numerical results, we can find that as the numerical ranges of experiment results are serious overlapped, CTM is no longer effective as indicator to classify or recognize the mentioned cardiovascular diseases. In figure (b) and table 1, for indicator D, considerable overlap exists between the three datasets, nsr2db, cudb and mitdb, which means that the quantitative indicator is not sufficiently suitable to distinguish the three physiological statuses of cardiovascular system effectively. The other figures are experiment results of CCTM$_k$ (k=1,2,3,4). Similarly, except for cudb, CCTMs for other four datasets have no significant difference and cannot be feasibly taken as feature to distinguish between the datasets.

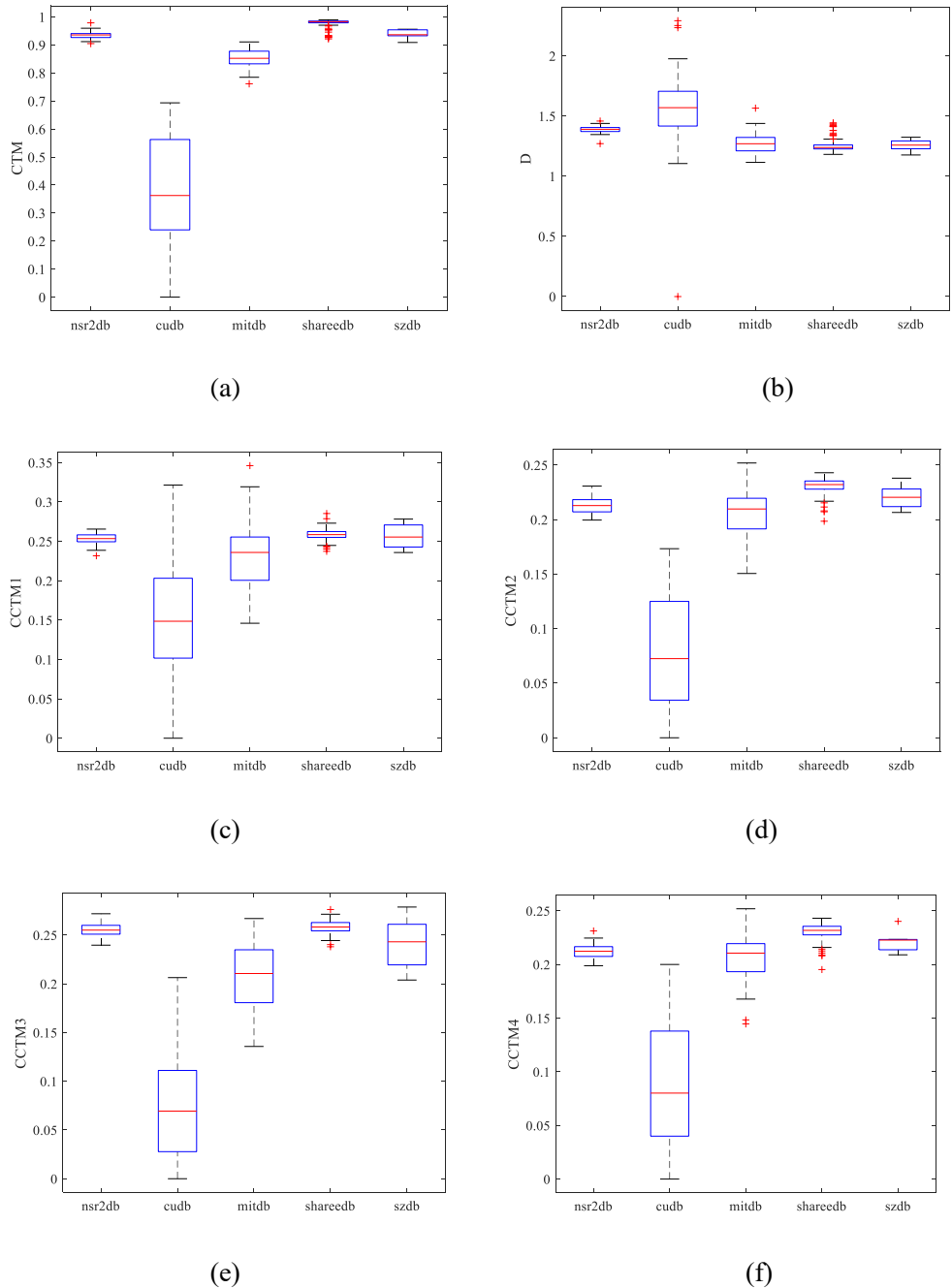

Figure 9. Boxplots for cudb, mitdb, nsr2db, shareedb and szdb. (a)~(f) are experiment results for the indicators CTM, D, CCTM$_1$, CCTM$_2$, CCTM$_3$ and CCTM$_4$ respectively.

**Table 1.** The means and standard deviations of CTM, D and CCTM.

| indicators | nsr2db | cudb | mitdb | shareedb | szdb |
|---|---|---|---|---|---|
| CTM | 0.93±0.01 | 0.39±0.20 | 0.85±0.03 | 0.98±0.01 | 0.94±0.02 |
| D | 1.39±0.03 | 1.57±0.39 | 1.28±0.09 | 1.25±0.05 | 1.26±0.05 |
| CCTM1 | 0.25±0.01 | 0.16±0.08 | 0.23±0.04 | 0.26±0.01 | 0.26±0.02 |
| CCTM2 | 0.21±0.01 | 0.08±0.05 | 0.21±0.02 | 0.23±0.01 | 0.22±0.01 |
| CCTM3 | 0.26±0.01 | 0.07±0.06 | 0.21±0.03 | 0.26±0.01 | 0.24±0.03 |
| CCTM4 | 0.21±0.01 | 0.08±0.06 | 0.21±0.02 | 0.23±0.01 | 0.22±0.01 |

The indicators are usually employed to quantify the variation of physiological status of the heart. Apparently, for the five datasets, the mentioned indicators CTM, CCTM and D cannot well meet the expectations of this study.

For TVM, experiment results of four quadrants are shown in Figure 10 and table 2. The boxplots of $E_{TV}$ show that the results of four quadrants are largely similar. For the four datasets nsr2db, cudb, mitdb and szdb, the numerical ranges almost have no overlap. For example, the means and standard deviations of $E_{TV}$ in the quadrant I are 118.57±35.39, 4.10±1.20, 6.78±2.41 and 16.11±5.58, respectively. The statistical results indicate that, as an indicator, $E_{TV}$ in quadrant I has the ability to distinguish the four physiological statuses of the heart and $E_{TV}$ as a kind of feature can be employed in the classification and recognition of the mentioned cardiovascular diseases. In the other quadrants, there are still great difference of $E_{TV}$ between the four datasets. Thus, the physiological features of aforementioned datasets can be extracted more effectively by $E_{TV}$ and experimental results indicate that the indicator has the consistency of performance in the four quadrants.

On the other hand, there is a certain overlap of numerical range of $E_{TV}$ between the box (interquartile range) of nsr2db and whisker (lower quartile) of chareedb. However, for the standard deviations of experiment results, there are significant difference between nsr2db and chareedb. Figure 10 and table 2 show that the stand deviation of chareedb is greater than nsr2db's, which means that the distribution density of experimental results of chareedb is partly lower than nsr2db's. Hence, the features of two datasets nsr2db and chareedb can be described effectively by $E_{TV}$. Via the indicator, two datasets can be still recognized.

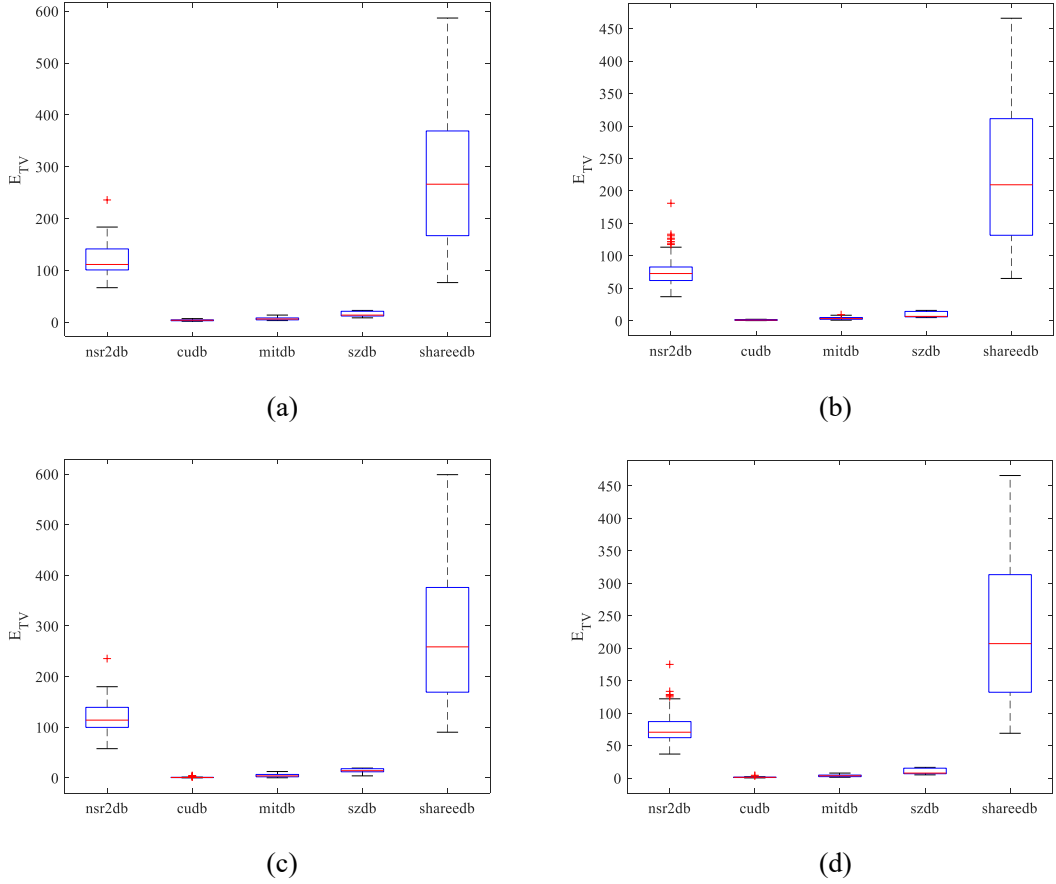

Figure 10. Boxplots of $E_{TV}$ for nsr2db, cudb, mitdb, szdb and shareedb. (a)~(d) are experiment results of $E_{TV}$ in quadrant I, II, II and IV, respectively.

**Table 2.** The means and standard deviations of $E_{TV}$ in the quadrant I, II, III and IV.

| quadrant | nsr2db | cudb | mitdb | szdb | shareedb |
|---|---|---|---|---|---|
| I | 118.57±35.39 | 4.10±1.20 | 6.78±2.41 | 16.11±5.58 | 277.13±120.81 |
| II | 78.29±29.63 | 1.10±0.54 | 3.80±1.98 | 9.52±4.72 | 227.19±103.56 |
| III | 118.30±36.02 | 1.15±1.04 | 4.63±3.05 | 13.83±5.22 | 279.79±124.76 |
| IV | 77.68±29.45 | 1.44±0.78 | 3.78±1.67 | 10.32±4.78 | 228.19±104.31 |

To adequately illustrate the performance of TVM further, here $E_{TV}$ and the other indicators are used for the classification of normal sinus rhythm and the other four cardiovascular diseases. Through the comparison of classification accuracy of $E_{TV}$ and the other indicators, the performance of TVM will be demonstrated intuitively. As a basic classification method, in this section, *k*-means algorithm is adopted to accomplish classification task. Meanwhile, *RI* is used to calculate the classification accuracy, which is defined as follows:

$$RI = \frac{CD}{TD} \qquad (12)$$

where *CD* and *TD* are the quantities of correct and total decisions, respectively.

Table 3 shows the classification results between the five datasets. Except the classification tasks of two datasets nsr2db and chareedb, the classification accuracies of $E_{TV}$ in four quadrants are significantly superior to the other indicators', which indicates that the physiological status of cardiovascular can be described by TVM more reasonably and precisely. For the classification of nsr2db and chareedb, the classification accuracy of $E_{TV}$ is lower than CTM and D, which means that CTM and D can better extract the feature of the dataset chareedb. However, as the significant difference of standard deviations for the two datasets, the classification accuracy of TVM should be improved further, if a density-based clustering algorithm is adopted in experiment.

**Table 3.** The clustering results of CTM, D, CCTM and $E_{TV}$.

| indicators | *RI* | | | | | | | | | |
|---|---|---|---|---|---|---|---|---|---|---|
| | nsr2db, cudb | nsr2db, mitdb | nsr2db, szdb | nsr2db, shareedb | cudb, mitdb | cudb, szdb | cudb, shareedb | mitdb, szdb | mitdb, shareedb | szdb, shareedb |
| CTM | 0.93 | 0.97 | 0.57 | **0.93** | 0.85 | 0.93 | 0.90 | **0.93** | 1 | 0.79 |
| D | 0.51 | 0.84 | 0.95 | **0.93** | 0.60 | 0.57 | 0.75 | 0.64 | 0.75 | 0.79 |
| $CCTM_I$ | 0.79 | 0.68 | 0.71 | 0.67 | 0.80 | 0.86 | 0.85 | 0.79 | 0.80 | 0.64 |
| $CCTM_{II}$ | 0.91 | 0.67 | 0.71 | 0.87 | 0.85 | **0.93** | 0.90 | 0.71 | 0.75 | 0.71 |
| $CCTM_{III}$ | 0.91 | 0.76 | 0.71 | 0.64 | 0.85 | **0.93** | 0.90 | 0.64 | 0.75 | 0.71 |
| $CCTM_{IV}$ | 0.87 | 0.67 | 0.71 | 0.87 | 0.85 | **0.93** | 0.85 | 0.64 | 0.75 | 0.86 |
| $TVM_I$ | **1** | **1** | **1** | 0.76 | **0.89** | 0.93 | **1** | 0.86 | 0.93 | **1** |
| $TVM_{II}$ | **1** | 0.99 | **1** | 0.76 | 0.86 | 0.86 | 0.95 | 0.79 | 0.95 | 0.93 |
| $TVM_{III}$ | **1** | 0.99 | **1** | 0.75 | 0.86 | 0.86 | **1** | **0.93** | **1** | 0.93 |
| $TVM_{IV}$ | **1** | 0.97 | **1** | 0.77 | 0.85 | 0.86 | **1** | 0.79 | **1** | 0.93 |

In this section, the performance of the temporal variation measure algorithm is preliminarily verified. By analyzing experiment results, TVM is suitable for properly describing the variation of RR interval time series and extracting the physiological status of cardiovascular system precisely.

# 5. Conclusion

As a development of Poincaré plot, the second-order difference plot is widely employed to depict the variation of physiological signal. In SODP, the different distribution pattern of scatter points represents various physiological status of the heart.

In the study, many quantitative indicators are introduced briefly, such as CTM, CCTM and D. However, the distribution pattern of points in SODP are sketchily described by these indicators and the abundant physiological information of dynamic change of cardiovascular system can not be adequately extracted. Therefore, the temporal variation measure algorithm is presented to remedy the limitations of the mentioned indicators partly.

In this paper, besides the study of distribution pattern of scatter points, the focus for research is twofold. Firstly, it is the change and span of time increment of successive RR intervals. Secondly, how can we extract the variation information of physiological status of the heart contained in a single point. Through the research, the difference of horizontal and vertical coordinates of point in SODP reflects the change of time increment effectively. The span of time increment can be properly depicted by Euclidean distance between the point and the origin. Then, the temporal variation measure algorithm is designed and the steps of algorithm is presented. Via TVM, the mentioned variation information is reconstructed in a three dimensional second-order difference plot. In the novel space, the variation information contained in the overall and local distribution pattern of all scatter points are well retained.

Experiment results preliminarily indicate that TVM is effective for extracting more physiological features from these scatter points in SODP. For the task of classification, the sinus rhythm can be effectively identified from the other physiological statuses via TVM and the accuracy of our method is higher than the others. Therefore, TVM can be as a valuable analytical method to applied in the physiological signal recognition and classification.

# 6. Acknowledgments

This work is supported by National Natural Science Foundation (NNSF) of China (Grant 61867005), and Natural Science Foundation of Ningxia Province (Grant 2020AAC03068). The authors gratefully acknowledge the insightful comments and suggestions from the reviewers and editor, which have helped to improve the presentation.

# 7. Conflict of Interest

The authors declare that there is no conflict of interest regarding the publication of this paper.